\documentclass[11pt]{article}

\usepackage{palatino}
\usepackage{mathpazo}
\usepackage{stmaryrd}

\usepackage{hyperref}
\hypersetup{pdfpagemode=UseNone}

\usepackage{amsfonts}
\usepackage{amssymb}
\usepackage{amsmath}
\usepackage{latexsym}
\usepackage{amsthm}
\usepackage{eepic}
\usepackage{sectsty}
\usepackage{graphicx}

\newtheorem{theorem}{Theorem}
\newtheorem{lemma}[theorem]{Lemma}

\theoremstyle{definition}

\newcommand{\tinyspace}{\mspace{1mu}}
\newcommand{\microspace}{\mspace{0.5mu}}
\newcommand{\op}[1]{\operatorname{#1}}

\newcommand{\norm}[1]{\left\lVert\tinyspace#1\tinyspace\right\rVert}

\newcommand{\tr}{\operatorname{Tr}}

\newcommand{\ip}[2]{\left\langle #1 , #2\right\rangle}

\def\({\left(}
\def\){\right)}
\def\I{\mathbb{1}}

\newcommand{\fid}{\operatorname{F}}
\newcommand{\setft}[1]{\mathrm{#1}}
\newcommand{\lin}[1]{\setft{L}\left(#1\right)}

\newcommand{\density}[1]{\setft{D}\left(#1\right)}

\newcommand{\herm}[1]{\setft{Herm}\left(#1\right)}
\newcommand{\pos}[1]{\setft{Pos}\left(#1\right)}
\newcommand{\channel}[1]{\setft{C}\left(#1\right)}
\newcommand{\pd}[1]{\setft{Pd}\left(#1\right)}

\def\real{\mathbb{R}}

\def \lket {\left|}
\def \rket {\right\rangle}
\def \lbra {\left\langle}
\def \rbra {\right|}
\newcommand{\ket}[1]{\lket\microspace #1 \microspace\rket}
\newcommand{\bra}[1]{\lbra\microspace #1 \microspace\rbra}

\newenvironment{mylist}[1]{\begin{list}{}{
	\setlength{\leftmargin}{#1}
	\setlength{\rightmargin}{0mm}
	\setlength{\labelsep}{2mm}
	\setlength{\labelwidth}{8mm}
	\setlength{\itemsep}{0mm}}}
	{\end{list}}

\newcommand{\reg}[1]{\mathsf{#1}}

\def\X{\mathcal{X}}
\def\Y{\mathcal{Y}}
\def\Z{\mathcal{Z}}

\begin{document}

\title{\bf Hedging bets with correlated quantum strategies}

\author{
  {\large Abel Molina \quad and \quad John Watrous}\\[2mm]
  {\it Institute for Quantum Computing and School of Computer Science}\\
  {\it University of Waterloo}}

\date{April 13, 2011}

\maketitle

\begin{abstract}
  This paper studies correlations among independently administered
  hypothetical tests of a simple interactive type, and demonstrates
  that correlations arising in quantum information theoretic variants
  of these tests can exhibit a striking non-classical behavior.
  When viewed in a game-theoretic setting, these correlations are
  suggestive of a perfect form of \emph{hedging}, where the risk of a
  loss in one game of chance is perfectly offset by one's actions in a
  second game.
  This type of perfect hedging is quantum in nature---it is not
  possible in classical variants of the tests we consider.
\end{abstract}

\section{Introduction}
\label{sec:introduction}

It is well known that quantum information theory allows for
correlations among measurement outcomes that are stronger than those
possible within any classical theory.
Bell inequality violations provide the archetypal example within this
category, where space-like separated measurements of entangled
particles yield correlated measurement outcomes that are incompatible
with local hidden-variable theories \cite{Bell64}.
This paper describes a different scenario in which this phenomenon
arises, and provides an example showing a striking difference between
quantum and classical theories in this scenario.

We will begin with an abstract description of the scenario we consider
that is mostly absent of precise discussions of underlying theories or
mathematical structures.
In simple terms, we imagine that one individual subjects another
individual to a \emph{test}, and for convenience we will refer to the
individual administering the test as \emph{Alice} and to the
test-taker as \emph{Bob}.
One may of course envision that Alice and Bob are devices rather than
individuals; we only choose the later point of view for the
convenience of using the names Alice and Bob.
The sort of tests under consideration are to have the following simple
form:
\begin{mylist}{\parindent}
\item[1.]
Alice prepares a \emph{question} and sends it to Bob.

\item[2.]
Bob responds by sending an \emph{answer} to Alice.

\item[3.]
Based on Bob's answer, as well as whatever memory she has of her own
question, Alice decides whether Bob has \emph{passed} or \emph{failed}
the test.
\end{mylist}

In a purely classical setting, one may imagine that Alice's behavior
is described by a probabilistic process, whereby her questions are
selected according to some probability distribution and her final decision
might also involve the use of randomness.
In the quantum setting, Alice's questions may take the form of quantum
information---possibly entangled with quantum memory of her own---and
she may expect quantum information from Bob in return.
In both the classical and quantum settings, we make the assumption
that Bob has a complete description of the process by which Alice
operates, and is generally interested in maximizing his probability of
passing the test.

For a fixed choice for Alice's test, let us let $p$ denote Bob's
\emph{optimal probability} of passing.
Formally speaking, without any assumptions on an underlying
mathematical model, $p$ may be defined to be the \emph{supremum} of all
passing probabilities for Bob, taken over all possible choices of his
strategy.
(In both the classical and quantum models, the supremum will always be
achieved, so that it may safely be replaced by the maximum.)
By assumption, Alice always makes a definitive decision about whether
Bob passes or fails, so he necessarily fails the test with probability
at least~$1-p$.

Now, consider that Alice instantiates two \emph{independent} copies of
her test: no correlations exist between the two questions that she
presents to Bob, and the processes by which she determines whether Bob
passes or fails are completely independent as well.
There are a variety of questions that one may ask about this type of
situation, including the following:
\begin{mylist}{\parindent}
\item[1.] What is the optimal probability with which Bob passes
  \emph{both} tests?
\item[2.] What is the optimal probability with which Bob passes 
  \emph{at least one} of the tests?
\end{mylist}

\noindent
It is natural to guess that Bob's optimal probability to pass both
tests is $p^2$, while his optimal probability to pass at least one
test is $1 - (1-p)^2$.
These are, of course, the optimal probabilities if he treats the two
tests independently.

In the classical setting, the probabilities $p^2$ and $1 - (1-p)^2$
are indeed optimal over all classical strategies, including those that
do not respect the independence of the two tests;
Bob cannot correlate the tests to his advantage in either case.
While these claims can be proved directly with little difficulty, we
will see that they fall out naturally as special cases in our
analysis of the quantum setting.

In the quantum setting, the natural guess is indeed correct for the
first question (as we will later discuss in greater detail): if Bob
aims to pass both tests, there is no advantage for him to correlate
the two tests.
This fact is known to those that have studied quantum interactive
proof systems \cite{KitaevW00}, and it is a consequence of a more
general result concerning semidefinite programs \cite{MittalS07}.
For the second question, on the other hand, the natural guess turns
out to be wrong.
We demonstrate this by giving an example where Bob can correlate
the two independent tests in such a way that he passes at least one of
the two tests \emph{with certainty}, despite the fact that $p<1$.
More specifically, our example describes a test where Bob's optimal
passing probability for a single instantiation of the test is
$\cos^2(\pi/8) \approx 0.85$, while he \emph{never} fails both tests
if he correlates two independent instantiations in the right way.

Bob's ability to correlate two independent tests in the way just
described can be seen as a perfect form of \emph{hedging}, as the
following (highly fictitious) scenario illustrates.
Bob is offered the opportunity to take part in two potentially lucrative but
somewhat risky games of chance, run by Alice.
The two games are completely independent and identical in nature: for
each he must put forth \$1 million of his own money to take part, and
he has 85\% chance to win if he plays optimally.
For each game he wins, Bob receives \$3 million (representing a \$2
million gain over his initial \$1~million investment), while he receives
nothing (and loses his \$1 million initial investment) if he loses.
For the sake of this example, we are to consider that a \$1 million or
greater loss means ruin for Bob.

These are, of course, highly compelling games of chance, and many
people would not hesitate to take out a \$2 million loan to play both:
the expected gain from each one is \$1,550,000, and the chance for
a loss in both, if they are treated independently, is only 2.25\%.
Bob, however, is a highly risk-averse person.
While he would enjoy being a millionaire, he cannot accept a 2.25\%
chance of ruin.
Classically speaking, Bob can do nothing to avoid at least a 2.25\%
chance of ruin, so he will choose not to play.
If the two games are modeled by quantum information as in our example,
however, Bob can be \emph{guaranteed} a \$1 million return, and
can therefore play without fear: an appropriately chosen quantum
strategy allows him to hedge his bets perfectly.

\section{Preliminaries}
\label{sec:preliminaries}

We assume the reader to be familiar with the basics of quantum
information theory, and suggest Nielsen and Chuang \cite{NielsenC00}
to those who are not.
The purpose of this section is to summarize some of the notation and
basic concepts we make use of, and to highlight a couple of concepts
that may be less familiar to some readers.

\subsection*{Basic notation, states, measurements and channels}

For any finite-dimensional complex Hilbert space $\X$ we write
$\lin{\X}$ to denote the set of linear operators acting on $\X$, 
we write $\herm{\X}$ to denote the set of Hermitian operators acting
on $\X$, we write $\pos{\X}$ to denote the set of positive
semidefinite operators acting on $\X$, we write $\pd{\X}$ to denote
the set of positive definite operators acting on $\X$, and we write
$\density{\X}$ to denote the set of density operators acting on $\X$.
For Hermitian operators $A,B\in\herm{\X}$ the notations $A\geq B$ and
$B\leq A$ indicate that $A - B$ is positive semidefinite, and the
notations $A > B$ and $B < A$ indicate that $A - B$ is positive definite.

Given operators $A,B\in\lin{\X}$, one defines the inner product
between $A$ and $B$ as $\ip{A}{B} = \tr(A^{\ast}B)$.
For Hermitian operators $A,B\in\herm{\X}$ it holds that
$\ip{A}{B}$ is a real number and satisfies $\ip{A}{B} = \ip{B}{A}$.
For every choice of finite-dimensional complex Hilbert space $\X$ and
$\Y$, and for a given linear mapping of the form
$\Phi:\lin{\X}\rightarrow\lin{\Y}$, there is a unique mapping
$\Phi^{\ast}:\lin{\Y}\rightarrow\lin{\X}$ (known as the \emph{adjoint}
of $\Phi$) that satisfies
$\ip{Y}{\Phi(X)} = \ip{\Phi^{\ast}(Y)}{X}$ for all $X\in\lin{\X}$ and
$Y\in\lin{\Y}$.

A \emph{register} is a hypothetical device that stores quantum
information.
Associated with a register $\reg{X}$ is a finite-dimensional complex
Hilbert space $\X$, and each quantum state of $\reg{X}$ is described
by a density operator $\rho\in\density{\X}$.
\emph{Qubits} are registers for which $\dim(\X) = 2$.
A \emph{measurement} of $\reg{X}$ is described by a set of positive
semidefinite operators $\{P_a\,:\,a\in\Sigma\}\subset\pos{\X}$,
indexed by a finite non-empty set of measurement outcomes $\Sigma$,
and satisfying the constraint $\sum_{a\in\Sigma}P_a = \I_{\X}$ (the
identity operator on $\X$).
If such a measurement is performed on $\reg{X}$ while it is in the
state $\rho$, each outcome $a\in\Sigma$ results with probability
$\ip{P_a}{\rho}$.
A \emph{quantum channel} is a completely positive and trace-preserving
linear mapping of the form \mbox{$\Phi:\lin{\X}\rightarrow\lin{\Y}$} that
describes a hypothetical physical process that transforms each state
$\rho$ of a register $\reg{X}$ into the state $\Phi(\rho)$ of another
register $\reg{Y}$.
The set of all channels of this form is denoted $\channel{\X,\Y}$.
The identity channel that does nothing to a register $\reg{X}$ is
denoted $\I_{\lin{\X}}$.

The Hilbert space corresponding to a pair of registers
$(\reg{X}_1,\reg{X}_2)$ is the tensor product $\X_1\otimes\X_2$ of the
spaces corresponding to $\reg{X}_1$ and $\reg{X}_2$.
Independent states, measurements and channels are represented by
elementary tensors in the following straightforward way:
\begin{mylist}{\parindent}
\item[1.]
  If registers $\reg{X}_1$ and $\reg{X}_2$ are independently prepared
  in states $\rho_1$ and $\rho_2$, then the state of the pair
  $(\reg{X}_1,\reg{X}_2)$ is given by the density operator
  $\rho_1\otimes\rho_2$.

\item[2.]
  If registers $\reg{X}_1$ and $\reg{X}_2$ are independently measured
  with respect to the measurements described by the collections
  $\{P_{a_1}\,:\,a_1\in\Sigma_1\}\subset\pos{\X_1}$ and
  $\{P_{a_2}\,:\,a_2\in\Sigma_2\}\subset\pos{\X_2}$, the resulting
  measurement on the pair $(\reg{X}_1,\reg{X}_2)$ is described by the
  collection $\{P_{(a_1,a_2)}\,:\,(a_1,a_2)\in\Sigma_1\times\Sigma_2\}$,
  where $P_{(a_1,a_2)} = P_{a_1}\otimes P_{a_2}\in\pos{\X_1\otimes\X_2}$.

\item[3.]
  If registers $\reg{X}_1$ and $\reg{X}_2$ are independently
  transformed into registers $\reg{Y}_1$ and $\reg{Y}_2$ according to
  the channels $\Phi_1\in\channel{\X_1,\Y_1}$ and
  $\Phi_2\in\channel{\X_2,\Y_2}$, respectively, then the
  transformation of the pair $(\reg{X}_1,\reg{X}_2)$ into the pair
  $(\reg{Y}_1,\reg{Y}_2)$ is described by the channel
  $\Phi_1\otimes\Phi_2\in\channel{\X_1\otimes\X_2,\Y_1\otimes\Y_2}$.
\end{mylist}

\subsection*{Linear mappings on operator spaces}

Suppose $\op{dim}(\X) = n$ and assume that a standard orthonormal basis
$\{\ket{1},\ldots,\ket{n}\}$ of $\X$ has been selected.
With respect to this basis, one defines the Choi-Jamio{\l}kowski
operator $J(\Phi)\in\lin{\Y\otimes\X}$ of a linear mapping
$\Phi:\lin{\X}\rightarrow\lin{\Y}$ as
\[
J(\Phi) = \sum_{1\leq i,j \leq n}
\Phi(\ket{i}\bra{j}) \otimes \ket{i}\bra{j}
\]
The mapping $J$ is a linear bijection from the space of mappings of
the form $\Phi:\lin{\X}\rightarrow\lin{\Y}$ to the operator space
$\lin{\Y\otimes\X}$.
It is well-known that $\Phi$ is completely positive if and only if
$J(\Phi) \in \pos{\Y\otimes\X}$, and that $\Phi$ is trace-preserving
if and only if $\tr_{\Y}(J(\Phi)) = \I_{\X}$
\cite{Choi75,Jamiolkowski72}.

While the Choi-Jamio{\l}kowski operator of a linear mapping
$\Phi:\lin{\X}\rightarrow\lin{\Y}$ is most commonly considered when
$\Phi$ is a channel, the concept is useful in more general settings
(as is illustrated in \cite{GutoskiW07} and \cite{ChiribellaDP09}, for
instance).
The following lemma, whose proof makes use of the Choi-Jamio{\l}kowski
operator of a particular mapping, gives one technical example that
will be useful later in the paper.

\begin{lemma} \label{lemma:Choi1}
  For every operator $A\in\lin{\X\otimes\Z}$ there exists a mapping
  $\Psi_A:\lin{\Z}\rightarrow\lin{\X}$ that possesses the following
  property:
  for every mapping $\Phi:\lin{\X}\rightarrow\lin{\Y}$ and every
  operator $B\in\lin{\Y\otimes\Z}$, the equation
  \[
  \ip{B}{\left(\Phi\otimes\I_{\lin{\Z}}\right)(A)}
  = \ip{\left(\I_{\lin{\Y}}\otimes\Psi_A\right)(B)}{J(\Phi)}
  \]
  holds.
  Moreover, if $A$ is positive semidefinite, then $\Psi_A$ is completely
  positive.
\end{lemma}

\begin{proof}
  The unique mapping $\Psi_A:\lin{\Z}\rightarrow\lin{\X}$ for which
  $J(\Psi) = \overline{A}$ (the entry-wise complex conjugate of $A$)
  possesses the required property.
  This fact is easily verified for operators $A$ taking the form
  $A = \ket{i}\bra{j} \otimes \ket{k}\bra{l}$, 
  in which case
  \[
  \Psi_A(Z) = \ket{i}\!\bra{k} \! Z \! \ket{l}\!\bra{j},
  \]
  and it follows for
  general operators by the conjugate-linearity/linearity of the inner
  product.
  Under the assumption that $A$ is positive semidefinite, so too is
  $\overline{A}$, from which it follows that $\Psi_A$ is completely
  positive.
\end{proof}

\subsection*{Semidefinite programming}

Semidefinite programming is a topic that has found several interesting
applications within quantum computing and quantum information theory
in recent years.
It is a valuable analytic tool, as well as a computational one.
Here, we provide just a brief summary of semidefinite programming that
is focused on the narrow aspects of it that we use.
More comprehensive discussions can be found in
\cite{VandenbergheB96,Lovasz03,deKlerk02,BoydV04}, for instance.

\pagebreak[3]

A semidefinite program is a triple $(\Phi,A,B)$, where
\begin{mylist}{\parindent}
\item[1.] 
$\Phi: \lin{\X} \rightarrow \lin{\Y}$ is a Hermiticity-preserving
  linear mapping, and
\item[2.] $A\in\herm{\X}$ and $B\in\herm{\Y}$ are Hermitian operators,
\end{mylist}
for some choice of finite-dimensional complex Hilbert spaces $\X$ and $\Y$.
We associate with the triple $(\Phi,A,B)$ two optimization problems,
called the \emph{primal} and \emph{dual} problems, as follows:
\begin{center}
  \begin{minipage}{2.6in}
    \centerline{\underline{Primal problem}}\vspace{-7mm}
    \begin{align*}
      \text{maximize:}\quad & \ip{A}{X}\\
      \text{subject to:}\quad & \Phi(X) = B,\\
      & X\in\pos{\X}.
    \end{align*}
  \end{minipage}
  \hspace*{13mm}
  \begin{minipage}{2.6in}
    \centerline{\underline{Dual problem}}\vspace{-7mm}
    \begin{align*}
      \text{minimize:}\quad & \ip{B}{Y}\\
      \text{subject to:}\quad & \Phi^{\ast}(Y) \geq A,\\
      & Y\in\herm{\Y}.
    \end{align*}
  \end{minipage}
\end{center}
\noindent
The optimal primal value of this semidefinite program is
\[
\alpha = \sup\{\ip{A}{X}\,:\,X\in\pos{\X},\,\Phi(X) = B\}
\]
and the optimal dual value is
\[
\beta = \inf\{\ip{B}{Y}\,:\,Y\in\herm{\Y},\,\Phi^{\ast}(Y) \geq A\}.
\]
(It is to be understood that the supremum over an empty set is
$-\infty$ and the infimum over an empty set is $\infty$, so $\alpha$
and $\beta$ are well-defined values in the set
$\real\cup\{-\infty,\infty\}$.
Our interest, however, will only be with semidefinite programs for
which $\alpha$ and $\beta$ are finite.)

It always holds that $\alpha \leq \beta$, which is a fact known as
\emph{weak duality}.
The condition $\alpha = \beta$, which is known as 
\emph{strong duality}, does not hold for every semidefinite program,
but there are simple conditions known under which it does hold.
The following theorem provides one such condition (that has both a
primal and dual form).

\begin{theorem}[Slater's theorem for semidefinite programs]
  \label{theorem:Slater}
Let $(\Phi,A,B)$ be a semidefinite program and let $\alpha$ and
$\beta$ be its optimal primal and dual values.
\begin{mylist}{\parindent}
\item[1.]
  If $\beta$ is finite and there exists a positive definite operator
  $X\in\pd{\X}$ for which $\Phi(X) = B$,
  then $\alpha = \beta$ and there exists an operator $Y\in\herm{\Y}$
  such that $\Phi^{\ast}(Y)\geq A$ and $\ip{B}{Y} = \beta$.
\item[2.]
  If $\alpha$ is finite and there exists a Hermitian operator
  $Y\in\herm{\Y}$ for which $\Phi^{\ast}(Y) > A$,
  then $\alpha = \beta$ and there exists a positive semidefinite 
  operator $X\in\pos{\X}$ such that $\Phi(X)=B$ and 
  $\ip{A}{X} = \alpha$.
\end{mylist}
\end{theorem}

In words, the first item of this theorem states that if the dual
problem is feasible and the primal problem is 
\emph{strictly feasible}, then strong duality holds and the optimal
dual solution is achievable.
The second item is similar, with the roles of the primal and dual
problems reversed.

\section{Interactive measurements}
\label{sec:interactive-measurements}

We now discuss the scenario described in the introduction in greater
mathematical detail, focusing on the quantum setting.
As is to be expected, the classical setting may be seen as a special
case of the quantum setting.

\begin{figure}[ht]
  \begin{center}
    \includegraphics{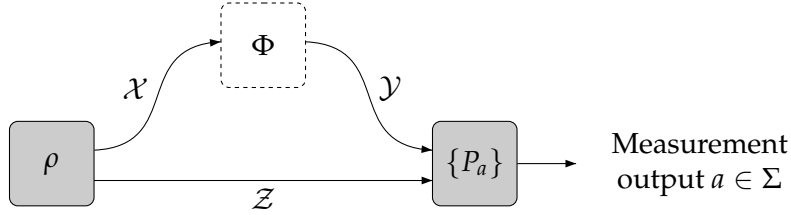}
  \end{center}
  \vspace{-2mm}
  \caption{An interactive measurement, consisting of the preparation
    of a state $\rho\in\density{\X\otimes\Z}$ followed by a measurement
    $\{P_a\,:\,a\in\Sigma\}\subset\pos{\Y\otimes\Z}$ on the space
    $\Y\otimes\Z$. 
    The interactive measurement is applied to a given quantum channel
    $\Phi\in\channel{\X,\Y}$, which is illustrated by a dashed box in
    the figure.}
  \label{fig:interactive-measurement}
\end{figure}

Tests of the form described in the introduction are modeled by
\emph{interactive measurements}, which are essentially measurements of
quantum channels:
an interactive measurement consists of a state preparation and a
measurement, to be applied to a given quantum channel.
More formally speaking, an interactive measurement is specified by
three finite-dimensional complex Hilbert spaces
$\X$, $\Y$ and $\Z$, along with two objects defined over these spaces:
\begin{mylist}{\parindent}
\item[1.]
  A state on the spaces $\X$ and $\Z$, represented by a density
  operator $\rho\in\density{\X\otimes\Z}$.
\item[2.]
  A measurement $\{P_a\,:\,a\in\Sigma\}\subset\pos{\Y\otimes\Z}$ on
  the spaces $\Y$ and $\Z$.
\end{mylist}
If such an interactive measurement is applied to a given channel
$\Phi\in\channel{\X,\Y}$, the probability associated with each
measurement outcome $a\in\Sigma$ is given by
\[
p(a) = \ip{P_a}{(\Phi\otimes\I_{\lin{\Z}})(\rho)}.
\]
An interactive measurement of a channel $\Phi$ is illustrated in 
Figure~\ref{fig:interactive-measurement}.

Suppose that an interactive measurement, specified by
a state $\rho\in\density{\X\otimes\Z}$ and a measurement 
$\{P_a\,:\,a\in\Sigma\}\subset\pos{\Y\otimes\Z}$ has been fixed.
For a given measurement outcome $a\in\Sigma$, one may consider both the
\emph{maximum} and \emph{minimum} probability with which the outcome
$a$ appears, over all choices of the quantum channel $\Phi$
upon which the interactive measurement is performed.
Let us denote the maximum probability by $M(a)$ and the minimum
probability by $m(a)$ for each $a\in\Sigma$, so that
\begin{align*}
M(a) & = \max_{\Phi\in\channel{\X,\Y}}
\ip{P_a}{(\Phi\otimes\I_{\lin{\Z}})(\rho)}\\
m(a) & = \min_{\Phi\in\channel{\X,\Y}}
\ip{P_a}{(\Phi\otimes\I_{\lin{\Z}})(\rho)}.
\end{align*}
(One notes that the above quantities are the maximization and
minimization, respectively, of a linear function on the compact set
of quantum channels $\channel{\X,\Y}$.
Thus, the use of the maximum and minimum rather than the supremum and
infimum are justified.)

The quantities $M(a)$ and $m(a)$ are expressible as the
optimal values of semidefinite programs, as we now describe.
For each $a\in\Sigma$ we let $Q_a\in\pos{\Y\otimes\X}$ be defined as
\begin{equation} \label{eq:Q-definition}
Q_a = \left(\I_{\lin{\Y}} \otimes \Psi_{\rho}\right)(P_a),
\end{equation}
for $\Psi_\rho$ being the mapping described by
Lemma~\ref{lemma:Choi1}.
We then have the following equality for each $a\in\Sigma$ and any
choice of a channel $\Phi\in\channel{\X,\Y}$:
\[
p(a) = \ip{P_a}{(\Phi\otimes\I_{\lin{\Z}})(\rho)}
=
\ip{Q_a}{J(\Phi)}.
\]
It therefore holds that
\[
M(a) = \max_{\Phi\in\channel{\X,\Y}} \ip{Q_a}{J(\Phi)}
\quad\quad\text{and}\quad\quad
m(a) = \min_{\Phi\in\channel{\X,\Y}} \ip{Q_a}{J(\Phi)}.
\]
The operator $J(\Phi)$ ranges over all choices of
$X\in\pos{\Y\otimes\X}$ satisfying $\tr_{\Y}(X)=\I_{\X}$ as $\Phi$ 
ranges over all channels $\Phi\in\channel{\X,\Y}$, and therefore the
following semidefinite program has optimal primal value $M(a)$:
\begin{center}
  \begin{minipage}{2.6in}
    \centerline{\underline{Primal problem}}\vspace{-7mm}
    \begin{align*}
      \text{maximize:}\quad & \ip{Q_a}{X}\\
      \text{subject to:}\quad & \tr_{\Y}(X) = \I_{\X},\\
      & X\in\pos{\Y\otimes\X}.
    \end{align*}
  \end{minipage}
  \hspace*{13mm}
  \begin{minipage}{2.6in}
    \centerline{\underline{Dual problem}}\vspace{-7mm}
    \begin{align*}
      \text{minimize:}\quad & \tr(Y)\\
      \text{subject to:}\quad & \I_{\Y} \otimes Y \geq Q_a,\\
      & Y\in\herm{\X}.
    \end{align*}
  \end{minipage}
\end{center}

\noindent
A slight modification yields a semidefinite program whose
optimal primal value is $m(a)$:
\begin{center}
  \begin{minipage}{2.6in}
    \centerline{\underline{Primal problem}}\vspace{-7mm}
    \begin{align*}
      \text{minimize:}\quad & \ip{Q_a}{X}\\
      \text{subject to:}\quad & \tr_{\Y}(X) = \I_{\X},\\
      & X\in\pos{\Y\otimes\X}.
    \end{align*}
  \end{minipage}
  \hspace*{13mm}
  \begin{minipage}{2.6in}
    \centerline{\underline{Dual problem}}\vspace{-7mm}
    \begin{align*}
      \text{maximize:}\quad & \tr(Y)\\
      \text{subject to:}\quad & \I_{\Y} \otimes Y \leq Q_a,\\
      & Y\in\herm{\X}.
    \end{align*}
  \end{minipage}
\end{center}
(The most straightforward way to fit this semidefinite program to the
precise formalism described in Section~\ref{sec:preliminaries} is to
exchange maximums and minimums and replace $Q_a$ with $-Q_a$, which
yields a semidefinite program for $-m(a)$.
One could alternately extend the definition of semidefinite programs
in a straightforward way to allow for minimizations in the primal
problem and maximizations in the dual.
The particular choice of these alternatives that one takes has no
effect on our analysis.)

It is clear that strict feasibility holds for each of the problems
presented: taking $X$ and $Y$ to be appropriately chosen scalar
multiples of the identity operator suffices to observe that these
properties hold.
Strong duality therefore holds for both semidefinite programs by
Theorem~\ref{theorem:Slater}, and optimal solutions are achieved for
each of the four problem formulations.

\section{Correlations among independent interactive measurements}

\begin{figure}[ht]
  \begin{center}
    \includegraphics{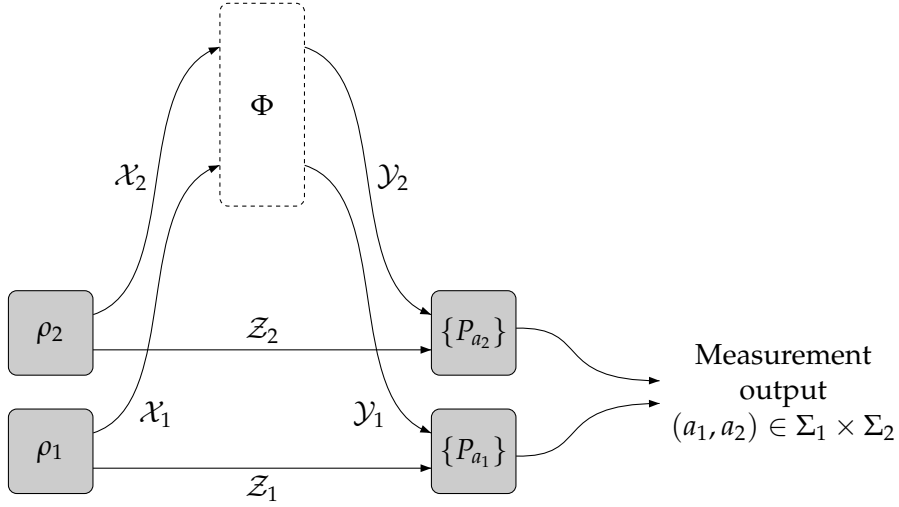}
  \end{center}
  \caption{Two interactive measurements, specified by pairs
    $(\rho_1,\{P_{a_1}\})$ and $(\rho_2,\{P_{a_2}\})$, are performed
    independently.
    The two interactive measurements are applied to a channel
    $\Phi\in\channel{\X_1\otimes\X_2,\Y_1\otimes\Y_2}$, which may not
    respect the independence of the two interactive measurements,
    potentially causing a correlation in the two measurement outcomes.}
  \label{fig:two-interactive-measurements}
\end{figure}

We now consider the situation in which two interactive measurements,
described by pairs\linebreak
$\left(\rho_1,\{P_{a_1}\,:\,a_1\in\Sigma_1\}\right)$ and
$\left(\rho_2,\{P_{a_2}\,:\,a_2\in\Sigma_2\}\right)$,
are performed independently, as suggested in
Figure~\ref{fig:two-interactive-measurements}.
%
While the interactive measurements are themselves performed
independently, it is not assumed that the channel
$\Phi\in\channel{\X_1\otimes\X_2,\Y_1\otimes\Y_2}$ respects this
independence.
Indeed, it is straightforward to devise examples where some choice of
the channel $\Phi$ causes a correlation in the outcomes produced by the
two measurements.
The main focus of this section is on the nature of the correlations
that are possible through the selection of various channels $\Phi$,
especially as these correlations relate to the scenario described in
Section~\ref{sec:introduction}.

Consider first the maximum output probability associated with a given
pair of measurement outcomes $(a_1,a_2)\in\Sigma_1\times\Sigma_2$.
In the following subsection we provide a proof that the maximum
probability $M(a_1,a_2)$ with which this pair is output is given by
\[
M(a_1,a_2) = M_1(a_1) \, M_2(a_2),
\]
where $M_1(a_1)$ and $M_2(a_2)$ denote the maximum output
probabilities of $a_1$ and $a_2$ with respect to the individual
interactive measurements with which they are associated.
Thus, to maximize the probability of outputting $(a_1,a_2)$, there is
absolutely no gain in choosing a channel $\Phi$ that correlates the
two interactive measurements: the optimal probability is achieved by
some choice $\Phi = \Phi_1\otimes\Phi_2$ that respects the independence
of the two interactive measurements.

Remarkably, a similar property does not generally hold when the
maximum is replaced by the minimum: we provide an example for which
$m_1(a_1) = m_2(a_2) = \sin^2(\pi/8) \approx 0.15$, but
$m(a_1,a_2) = 0$.

\subsection*{Analysis for multiplicativity}

To see that $M(a_1,a_2) = M_1(a_1) M_2(a_2)$, we may consider the
semidefinite program representing the optimal probability $M(a_1,a_2)$:
\begin{center}
  \begin{minipage}{3in}
    \centerline{\underline{Primal problem}}\vspace{-7mm}
    \begin{align*}
      \text{maximize:}\quad & \ip{Q_{a_1}\otimes Q_{a_2}}{X}\\
      \text{subject to:}\quad & \tr_{\Y_1\otimes\Y_2}(X) =
      \I_{\X_1\otimes\X_2},\\
      & X\in\pos{\Y_1\otimes\X_1\otimes\Y_2\otimes\X_2}.
    \end{align*}
  \end{minipage}
  \hspace*{3mm}
  \begin{minipage}{3.2in}
    \centerline{\underline{Dual problem}}\vspace{-7mm}
    \begin{align*}
      \text{minimize:}\quad & \tr(Y)\\
      \text{subject to:}\quad & \pi(\I_{\Y_1} \otimes \I_{\Y_2} \otimes
      Y)\pi^{\ast} \geq Q_{a_1}\otimes Q_{a_2},\\
      & Y\in\herm{\X_1\otimes\X_2}.
    \end{align*}
  \end{minipage}
\end{center}
(The unitary operator $\pi$ is defined by the action
$\pi(y_1 \otimes y_2 \otimes x_1 \otimes x_2) = 
y_1 \otimes x_1 \otimes y_2 \otimes x_2$ for all
$y_1\in\Y_1$, $y_2\in\Y_2$, $x_1\in\X_1$, $x_2\in\X_2$.)

One first observes that the inequality 
$M(a_1,a_2) \geq M_1(a_1) M_2(a_2)$ is straightforward:
the choice $X = X_1\otimes X_2$ for primal-optimal choices of $X_1$
and $X_2$ gives a primal feasible solution achieving the objective
value $M_1(a_1) M_2(a_2)$.

Similarly, the upper bound $M(a_1,a_2) \leq M_1(a_1) M_2(a_2)$
may be established by considering the dual problem.
For $Y_1\in\herm{\X_1}$ and $Y_2\in\herm{\X_2}$ being dual-optimal we
have $\tr(Y_1) = M(a_1)$ and $\tr(Y_2) = M(a_2)$, and thus
$\tr(Y_1\otimes Y_2) = M_1(a_1) M_2(a_2)$.
Moreover, as $Q_{a_1}$ and $Q_{a_2}$ are positive semidefinite and
the constrains $\I_{\Y_1} \otimes Y_1 \geq Q_{a_1}$ and
$\I_{\Y_2} \otimes Y_2 \geq Q_{a_2}$ hold, it follows that
$\I_{\Y_1}\otimes Y_1$ and $\I_{\Y_2} \otimes Y_2$ are positive
semidefinite.
Using the fact that $A\geq B$ and $C\geq D$ implies $A\otimes C \geq
B\otimes D$ for any choice of positive semidefinite operators $A$,
$B$, $C$ and $D$, we have
\[
\pi(\I_{\Y_1} \otimes \I_{\Y_2} \otimes Y_1 \otimes Y_2)\pi^{-1} 
=
(\I_{\Y_1} \otimes Y_1) \otimes (\I_{\Y_2} \otimes Y_2)
\geq
Q_{a_1}\otimes Q_{a_2}.
\]
The operator $Y_1\otimes Y_2$ is therefore dual feasible, so it is
established that $M(a_1,a_2) \leq M(a_1) M(a_2)$.

We note that this is a particular instance of a semidefinite program
obeying the \emph{product rule} considered by Mittal and Szegedy
\cite{MittalS07}, where the argument just presented is applied to a
more general class of semidefinite programs.

When the maximum is replaced by the minimum, however, the above
argument breaks down.
In this case, the semidefinite program whose optimal value is
$m(a_1,a_2)$ takes the following form:
\begin{center}
  \begin{minipage}{3in}
    \centerline{\underline{Primal problem}}\vspace{-7mm}
    \begin{align*}
      \text{minimize:}\quad & \ip{Q_{a_1}\otimes Q_{a_2}}{X}\\
      \text{subject to:}\quad & \tr_{\Y_1\otimes\Y_2}(X) =
      \I_{\X_1\otimes\X_2},\\
      & X\in\pos{\Y_1\otimes\X_1\otimes\Y_2\otimes\X_2}.
    \end{align*}
  \end{minipage}
  \hspace*{3mm}
  \begin{minipage}{3.2in}
    \centerline{\underline{Dual problem}}\vspace{-7mm}
    \begin{align*}
      \text{maximize:}\quad & \tr(Y)\\
      \text{subject to:}\quad & \pi(\I_{\Y_1} \otimes \I_{\Y_2} \otimes
      Y)\pi^{\ast} \leq Q_{a_1}\otimes Q_{a_2},\\
      & Y\in\herm{\X_1\otimes\X_2}.
    \end{align*}
  \end{minipage}
\end{center}

The upper-bound $m(a_1,a_2) \leq m_1(a_1) \, m_2(a_2)$ is easily
established by once again taking $X = X_1 \otimes X_2$ for primal
optimal points $X_1$ and $X_2$.
For the lower-bound
\[
m(a_1,a_2) \stackrel{?}{\geq} m_1(a_1)\,m_2(a_2),
\]
however, a problem arises: unlike the situation for the maximum, one
may not conclude that the operators $\I_{\Y_1}\otimes Y_1$ and
$\I_{\Y_2}\otimes Y_2$ are positive semidefinite for optimal dual
solutions $Y_1$ and $Y_2$ (and indeed they may not be positive
semidefinite in some cases).
One may fail to prove that the operator $Y = Y_1\otimes Y_2$ is
dual-feasible in this case, so that a lower-bound is not established.

\subsection*{An example showing non-classical behavior}

We now present our example of a quantum test that allows for a strong
correlation of the sort described in Section~\ref{sec:introduction}.
The test is as follows:
\begin{mylist}{\parindent}
\item[1.]
Alice prepares a pair of qubits $(\reg{X},\reg{Z})$ in the state
\[
u = \frac{1}{\sqrt{2}} \ket{00} + \frac{1}{\sqrt{2}} \ket{11}
\in\X\otimes\Z,
\]
and sends $\reg{X}$ to Bob.

\item[2.]
Bob applies any quantum channel he likes to $\reg{X}$, obtaining a
qubit $\reg{Y}$ that he sends back to Alice.
As a result of his action, the pair $(\reg{Y},\reg{Z})$ then has some
particular state $\sigma\in\density{\Y\otimes\Z}$.

\item[3.]
Alice measures $(\reg{Y},\reg{Z})$ with respect to the projective
measurement $\{P_0,P_1\}$, where $P_0 = \I - P_1$ and
$P_1 = v v^{\ast}$ for
\[
v = \cos(\pi/8) \ket{00} + \sin(\pi/8) \ket{11}.
\]
The outcome 1 is to be interpreted that Bob passes the test, while the
outcome 0 means that he fails.
\end{mylist}

Now, if Bob can produce a given state
$\sigma\in\density{\Y\otimes\Z}$ in step 2, it must hold that
\begin{equation} \label{eq:1}
\tr_{\Y}(\sigma) = 
\tr_{\X}(u u^{\ast}) = \frac{1}{2} \I_{\Z};
\end{equation}
no action that Bob performs on his registers can influence the
state of Alice's register.
The probability that Alice obtains the outcome 1 is
\[
\ip{P_1}{\sigma} = \fid(v v^{\ast},\sigma)^2,
\]
where $\fid(\cdot,\cdot)$ denotes the \emph{fidelity} function and
where the equality holds by virtue of the fact that $v v^{\ast}$ is
pure.
By the monotonicity of the fidelity function under partial tracing, we
have
\[
\fid(v v^{\ast},\sigma)^2 \leq \fid\left( \tr_{\Y}(v v^{\ast}),
\tr_{\Y}(\sigma)\right)^2
= \fid(Q,R)^2
\]
for
\[
Q = 
\begin{pmatrix}
  \cos^2(\pi/8) & 0 \\ 0 & \sin^2(\pi/8)
\end{pmatrix}
\quad\quad\text{and}\quad\quad
R = 
\begin{pmatrix}
  \frac{1}{2} & 0 \\ 0 & \frac{1}{2}
\end{pmatrix}.
\]
By a direct calculation we determine
\[
\fid(Q,R)^2 = \norm{\sqrt{Q}\sqrt{R}}_1^2 =
\frac{1}{2}\left(\cos(\pi/8) + \sin(\pi/8)\right)^2
= \cos^2(\pi/8).
\]
Alice therefore outputs 1, indicating that Bob passes the test,
with probability at most $\cos^2(\pi/8) \approx 0.85$.

Finally, for two instantiations of the test described above, we
consider what happens when Bob applies the phase flip
$\ket{00} \mapsto -\ket{00}$,
$\ket{01} \mapsto \ket{01}$,
$\ket{10} \mapsto \ket{10}$,
$\ket{11} \mapsto \ket{11}$
on the two qubits he receives.
Alice has prepared the state
\[
\frac{1}{2}\ket{0000} +
\frac{1}{2}\ket{0011} +
\frac{1}{2}\ket{1100} +
\frac{1}{2}\ket{1111}
\in \X_1\otimes\Z_1 \otimes \X_2 \otimes\Z_2,
\]
and Bob's phase flip transforms this state to
\[
-\frac{1}{2}\ket{0000} +
\frac{1}{2}\ket{0011} +
\frac{1}{2}\ket{1100} +
\frac{1}{2}\ket{1111}
\in \Y_1\otimes\Z_1 \otimes \Y_2 \otimes\Z_2.
\]
Writing
\[
w = - \sin(\pi/8)\ket{00} + \cos(\pi/8)\ket{11}
\]
we find that
\[
-\frac{1}{2}\ket{0000} +
\frac{1}{2}\ket{0011} +
\frac{1}{2}\ket{1100} +
\frac{1}{2}\ket{1111}
=
\frac{1}{\sqrt{2}}v\otimes w
+
\frac{1}{\sqrt{2}}w\otimes v.
\]
When Alice measures this state with respect to the measurement
$\{\Pi_0,\Pi_1\}$, she obtains exactly one outcome 0 and one
outcome~1.
Thus, it holds that $m(0,0) = 0$; Bob passes exactly one of the two
tests with certainty.

\subsection*{Analysis for the classical setting}

We now observe that the behavior exhibited in the example just
described cannot happen in the classical setting.

Suppose that $\rho\in\density{\X\otimes\Z}$ and
$\{P_a\,:\,a\in\Sigma\}\subset\pos{\Y\otimes\Z}$ describe an
interactive measurement as before.
As is typical in quantum information theory, the classical setting
corresponds to the special case in which these operators are all
diagonal (with respect to the standard basis).
Note that when the density operator $\rho$ and a given measurement
operator $P_a$ are diagonal, it holds that the operator
$Q_a\in\pos{\Y\otimes\X}$ defined by \eqref{eq:Q-definition} is also
diagonal.

Now suppose that $a\in\Sigma$ is a measurement outcome for which $Q_a$
is diagonal, and consider the semidefinite program whose optimal
primal value describes the minimum probability associated with the
outcome $a$ (i.e., whose optimal value is $m(a)$):
\begin{center}
  \begin{minipage}{2.6in}
    \centerline{\underline{Primal problem}}\vspace{-7mm}
    \begin{align*}
      \text{minimize:}\quad & \ip{Q_a}{X}\\
      \text{subject to:}\quad & \tr_{\Y}(X) = \I_{\X},\\
      & X\in\pos{\Y\otimes\X}.
    \end{align*}
  \end{minipage}
  \hspace*{13mm}
  \begin{minipage}{2.6in}
    \centerline{\underline{Dual problem}}\vspace{-7mm}
    \begin{align*}
      \text{maximize:}\quad & \tr(Y)\\
      \text{subject to:}\quad & \I_{\Y} \otimes Y \leq Q_a,\\
      & Y\in\herm{\X}.
    \end{align*}
  \end{minipage}
\end{center}
We will observe that there exists a dual optimal solution $Y$ that is
positive semidefinite.
It will be helpful for this purpose to let
$\Lambda_{\X}\in\channel{\X,\X}$ and $\Lambda_{\Y}\in\channel{\Y,\Y}$
denote the completely dephasing channels corresponding to $\X$ and
$\Y$, respectively.
More precisely, the mapping $\Lambda_{\X}$ is defined by
\[
\Lambda_{\X}(\ket{i}\!\bra{j}) = \left\{
\begin{array}{ll}
  \ket{i}\!\bra{j} & \text{if $i = j$}\\
  0 & \text{if $i\not=j$}
\end{array}
\right.
\]
for $1\leq i,j\leq n$, where $\{\ket{1},\ldots,\ket{n}\}$ is the
standard basis of $\X$, and $\Lambda_{\Y}$ is defined similarly with
respect to the standard basis of $\Y$.

Now consider an arbitrary dual optimal solution $Y_0\in\herm{\X}$.
The mapping $\Lambda_{\Y}\otimes\Lambda_{\X}$ is completely positive, so
the relation $\I_{\Y}\otimes Y_0 \leq Q_a$ implies that
\[
\I_{\Y} \otimes \Lambda_{\X}(Y_0) =
(\Lambda_{\Y}\otimes\Lambda_{\X})(\I_{\Y}\otimes Y_0)
\leq (\Lambda_{\Y}\otimes\Lambda_{\X})(Q_a)
= Q_a.
\]
The diagonal operator $\Lambda_{\X}(Y_0)$ is therefore dual feasible.
As $\Lambda_{\X}$ preserves trace, $\Lambda_{\X}(Y_0)$ achieves the same
dual objective value as $Y_0$, and is therefore optimal as well.
Finally, define
\[
Y = \sum_{i = 1}^n \max\{0,\,\bra{i}\Lambda_{\X}(Y_0)\ket{i}\}
\ket{i}\!\bra{i}.
\]
In other words, $Y$ is obtained from $\Lambda_{\X}(Y_0)$ by replacing
each negative diagonal entry with 0.
The inequality $\I_{\Y}\otimes Y \leq Q_a$ follows from the
inequality $\I_{\Y} \otimes \Lambda_{\X}(Y_0) \leq Q_a$ together
with the observation that each diagonal entry of $Q_a$ is necessarily
nonnegative (because $Q_a$ is positive semidefinite).
As $\tr(Y)\geq \tr(\Lambda_{\X}(Y_0)) = \tr(Y_0)$, it follows that
$Y$ is also dual optimal.
(The reality, of course, is that $Y = \Lambda_{\X}(Y_0)$, for
otherwise $\Lambda_{\X}(Y_0)$ would not have been dual optimal.)

Finally, consider the situation in which two classical interactive
measurements, described by pairs
$\left(\rho_1,\{P_{a_1}\,:\,a_1\in\Sigma_1\}\right)$ and
$\left(\rho_2,\{P_{a_2}\,:\,a_2\in\Sigma_2\}\right)$,
are performed.
One finds that the equality
\begin{equation} \label{eq:classical-min-multiplicative}
m(a_1,a_2) = m_1(a_1)\,m_2(a_2)
\end{equation}
considered before must now hold by an analysis similar to the one for
the maximum output probability case: positive semidefinite optimal
dual solutions exist for the semidefinite program described above for
each operator $Q_{a_1}$ and $Q_{a_2}$, allowing for the
straightforward construction of optimal primal and dual solutions
to the semidefinite program whose optimal value is $m(a_1,a_2)$,
thereby implying \eqref{eq:classical-min-multiplicative}.

\section{Conclusion}

This paper has considered correlated strategies against independently
administered hypothetical tests of a simple interactive type.
It has been demonstrated that correlations arising in quantum
information theoretic variants of these tests can exhibit a
non-classical \emph{hedging} type of behavior.

One may, of course, consider situations in which more than two
independent tests are performed, where a variety of statistics may be
of interest.
For example, one may consider Bob's optimal probability to pass some
threshold number $t$ of some (possibly large) number $k$ of
independently administered tests.
Based on our results we know that a surprising behavior exists even
for the case $t=1$ and $k=2$, and it would be interesting to
investigate the possible asymptotic behaviors that can arise.

The work of this paper is motivated by the problem of error reduction
through \emph{parallel repetition} for quantum interactive proof
systems.
In complexity theory, hypothetical tests along the lines of those we
have considered are often studied as a tool to classify computational
problems, and the resulting model is known as the 
\emph{interactive proof system} model \cite{GoldwasserMR89,BabaiM88}.
Interactive proof systems that allow for interactions consisting of
multiple rounds are often considered, but for the sake of this
discussion we will focus only on those interactive proof systems that
consist of a single question followed by a response---or, in other
words, those interactions that correspond to interactive measurements
as we have considered them in this paper.

In the context of interactive proof systems, the individual we have
called Alice is called the \emph{verifier} and Bob is called the
\emph{prover}.
The verifier's computational ability is limited (usually to
probabilistic or quantum polynomial time) while the prover's
computational ability is unrestricted.
For each input string $x$ to a fixed decision problem $L$, the prover
and verifier engage in an interaction wherein the prover attempts to
convince (or prove to) the verifier that the string $x$ should be
accepted as a yes-instance of the problem $L$.
To say that such a system is valid for the problem $L$ means two
things: one is that it must be possible for a prover to convince the
verifier to accept with high probability if the input is truly a
yes-instance of the problem, and the second is that the verifier must
reject no-instances of the problem with high probability regardless of
the prover's actions.
The first requirement is called the \emph{completeness} condition, and
is analogous to the condition in formal logic that true statements can
be proved.
The second condition is called the \emph{soundness} condition, and is
analogous to the condition that false statements cannot be proved.

Suppose now that a particular verifier has been specified (for a fixed
decision problem $L$) so that the following conditions hold:
\begin{mylist}{\parindent}
\item[1.]
  For each yes-instance $x$ to $L$, it is possible for a prover to
  convince the verifier to accept with probability at least $\alpha$.
\item[2.]
  For each no-instance $x$ to $L$, the verifier always rejects with
  probability at most $\beta$, regardless of the prover's actions.
\end{mylist}
It may be, for instance, that $\alpha = 1/2 + \delta$ and 
$\beta = 1/2 - \delta$ for some small constant $\delta > 0$.
A more desirable situation is one in which $\alpha$ is replaced by $1
- \varepsilon$ and $\beta$ is replaced by $\varepsilon$ for a small
value of~$\varepsilon$.
The process of specifying a new verifier based on the original one
that meets stronger completeness and soundness conditions, such as the
ones just suggested, is called \emph{error reduction}.

In a purely algorithmic situation, the natural way to reduce error is
to gather statistics from multiple independent executions of a given
algorithm.
For instance, if an algorithm outputs a binary value that is correct
(for worst-case inputs) with a probability of at least 2/3 on any
single execution of the algorithm, it is straightforward to obtain a
new algorithm with a very high probability of correctness: one simply
runs the original algorithm independently many times and takes the
majority value as the output.
A natural adaptation of this idea to interactive proof systems is to
define a new verifier that independently runs many instances of the
test performed by the original verifier, and accepts if and only if
some suitably chosen threshold number of these independent tests would
have led the original verifier to acceptance.
In the situation under consideration, one is to understand that it is
important for the new verifier to run these independent tests in
parallel (as opposed to requiring the prover to respond sequentially
to the individual tests).

It is not obvious that this works in the context of interactive proof
systems for precisely the reason that has been considered in this
paper: a hypothetical prover that interacts with many independent
executions of an interactive proof system need not respect the
independence of these executions.
Nevertheless, in the classical setting it has long been known 
that error reduction through parallel repetition followed by a
threshold value computation works perfectly\footnote{%
  The situation is very different for \emph{multi-prover} interactive
  proof systems, wherein the subject of parallel repetition is
  complicated \cite{Raz98,Holenstein09,Raz08}.}
for (single-prover) interactive proof systems.
To say that the reduction is perfect means that if $p$ is the optimal
success probability for the original verifier, then the optimal
probability to cause at least $t$ acceptances among $k$ independent
executions of the original verifier is
\begin{equation} \label{eq:binomial-sum}
\sum_{j = t}^k \binom{k}{j} p^j (1 - p)^{k-j}.
\end{equation}
In other words, a prover gains absolutely no advantage in trying to
correlate the independent tests performed by the verifier.

In the quantum setting, however, it was not previously known if
parallel repetition followed by a threshold value computation could
allow for a perfect error reduction (or indeed any error reduction at
all for certain values of $\alpha$ and $\beta$).
Our results show that parallel repetition followed by a threshold
value computation does not lead to a perfect reduction of error:
substituting $k = 2$, $t=1$ and $p = \cos^2(\pi/8)$ into
\eqref{eq:binomial-sum} yields an upper bound of approximately
$0.98$, which is violated by the strategy we described in the previous
section (which achieves the value 1).
We note that parallel repetition does work in the case of
\emph{perfect completeness} (i.e., $\alpha = 1$), wherein the
threshold value computation is replaced by the logical-and
\cite{KitaevW00}, and that there is a more complicated method for
error reduction (based on a logical-and of majorities), which does
allow for error reduction in the general case of the setting under
consideration \cite{JainUW09}.

Based on the semidefinite programming formalism we have described, it
is possible to prove an upper bound of
\[
\sum_{j = t}^k \binom{k}{j} p^j
\]
on the probability for a quantum prover to cause at least $t$
acceptances among $k$ independent executions as considered above.
Unfortunately this expression does not lead to a reduction of errors
for a wide range of choices of $\alpha > \beta$.
This bound yields a value larger than 1 in some situations, and when
the value is smaller than 1 we do not know how closely it can be
approached by a valid quantum strategy.


\subsection*{Acknowledgments}

Abel Molina acknowledges support from QuantumWorks, MITACS, a Mike
and Ophelia Lazaridis Graduate Fellowship and a David R. Cheriton
Graduate Scholarship.
John Watrous acknowledges support from NSERC, CIFAR, QuantumWorks and
MITACS.

\bibliographystyle{alpha}
\bibliography{hedging}

\end{document}